\documentclass{ws-procs9x6}

\def\to{\rightarrow}

\def\ve{\varepsilon}

\def\th{\theta}

\def\ka{\kappa}

\def\rh{\rho}

\def\si{\sigma}

\def\cl{{\mathcal L}}

\def\fr#1#2{{{#1} \over {#2}}}

\def\quar{{\textstyle{1\over 4}}}
\def\frac#1#2{{\textstyle{{#1}\over {#2}}}}

\def\lsim{\mathrel{\rlap{\lower4pt\hbox{\hskip1pt$\sim$}}
    \raise1pt\hbox{$<$}}}
\def\gsim{\mathrel{\rlap{\lower4pt\hbox{\hskip1pt$\sim$}}
    \raise1pt\hbox{$>$}}}
\def\sqr#1#2{{\vcenter{\vbox{\hrule height.#2pt
         \hbox{\vrule width.#2pt height#1pt \kern#1pt
         \vrule width.#2pt}
         \hrule height.#2pt}}}}

\def\prt{\partial}
\def\lrpartial{\raise 1pt\hbox{$\stackrel\leftrightarrow\partial$}}

\def\etal{{\it et al.}}

\newcommand{\beq}{\begin{equation}}
\newcommand{\eeq}{\end{equation}}
\newcommand{\bea}{\begin{eqnarray}}
\newcommand{\eea}{\end{eqnarray}}
\newcommand{\rf}[1]{(\ref{#1})}

\begin{document}

\title{Theoretical topics in spacetime-symmetry violations}

\author{R.\ LEHNERT}

\address{Center for Theoretical Physics\\
Massachusetts Institute of Technology\\
Cambridge, MA 02139, USA\\ 
E-mail: rlehnert@lns.mit.edu}

\maketitle

\abstracts{
The Lorentz- and CPT-violating Chern--Simons extension of electrodynamics 
is considered. 
In the context of $N=4$ supergravity in four spacetime dimensions, 
it is argued 
that cosmological solutions can generate this extension. 
Within Chern--Simons electrodynamics, 
theoretical and phenomenological topics are reviewed
that concern the number of the remaining spacetime symmetries 
and the vacuum Cherenkov effect, respectively. 
}

\section{Introduction}
\label{intro}

Spacetime-symmetry investigations 
owe their present popularity to the idea 
that Lorentz- and CPT-violation 
could be a signature from unknown physics 
possibly arising at the Planck scale. 
Research in this field 
may therefore be divided into three broad and overlapping areas: 
the identification of mechanisms 
that can generate Lorentz and CPT breakdown in underlying physics, 
the determination and theoretical study of the ensuing low-energy effects, 
and the low-energy phenomenology together with the corresponding experimental tests. 

During the last two decades, 
a number of theoretical arguments 
suggesting the possibility of spacetime-symmetry breaking 
in underlying physics
have been put forward. 
Examples of such arguments involve  
string field theory,\cite{ksp} 
realistic noncommutative field theories,\cite{ncqed} 
spacetime-varying fields,\cite{spacetimevarying} 
various quantum-gravity models,\cite{qg} 
nontrivial spacetime topology,\cite{klink} 
random-dynamics models,\cite{fn02} 
multiverses,\cite{bj} 
and brane-world scenarios.\cite{brane} 
Although the underlying dynamics remains Lorentz invariant in most of the above situations, 
Lorentz and CPT symmetry are nevertheless violated 
in the ground state at low energies. 
These ideas 
provide one of key motivations 
for Lorentz- and CPT-violation research.

At presently attainable energies, 
the effects resulting from Lorentz and CPT breakdown 
in underlying physics
are described by the Standard-Model Extension (SME)---an 
effective-field-theory framework 
containing the usual Standard Model\cite{flatsme} 
and general relativity.\cite{curvedsme} 
Various theoretical investigations 
and consistency analyses 
have been performed 
within the context of the SME.\cite{investigations,hariton07} 
While some of these studies 
have clarified conceptual issues,  
non have
suggested any internal inconsistencies.

The SME has also provided the basis 
for numerous phenomenological and experimental 
investigations of Lorentz and CPT violation.\cite{review} 
Specific analyses include, 
for example, 
ones with photons,\cite{randomphotonexpt,cherenkov} 
neutrinos,\cite{randomnuexpt}
electrons,\cite{randomeexpt} 
protons and neutrons,\cite{randompnexpt} 
mesons,\cite{randomhadronexpt} 
and muons.\cite{muexpt}  
These studies 
have placed tight constraints 
on numerous SME coefficients or combinations of them. 
Some of the obtained bounds 
can be considered to probe the Planck scale.

The above remarks demonstrate 
that all three of the aforementioned areas 
of Lorentz- and CPT-violation research 
are active and vibrant fields of scientific inquiry
spanning many physics disciplines. 
This talk aims at illustrating 
within a specific example---namely 
the Maxwell--Chern--Simons (MCS) model\cite{mcs} contained in the SME---how 
the three subfields are interwoven. 
Section \ref{sugra} shows  
that the Lorentz- and CPT-violating MCS model 
can arise in underlying Lorentz-invariant physics, 
more specifically in a low-energy cosmological context 
of $N=4$ supergravity. 
In Sec.\ \ref{symmetries}, 
some theoretical issues regarding the counting of symmetries in the MCS model 
are discussed at the level of the SME. 
A review of vacuum Cherenkov radiation, 
which is a phenomenological effect occuring in MCS theory, 
is presented in Sec.\ \ref{cherenkov}

\section{Emergence of the MCS model in supergravity cosmology}
\label{sugra}

The discussion in this section 
is based upon results obtained 
in the first paper of Ref.\ \refcite{spacetimevarying},
which considers pure $N=4$ supergravity
in four spacetime dimensions.
Although unrealistic in detail,
it is a limit of $N=1$ supergravity 
in eleven dimensions,
which is contained in M-theory.
We may thus expect
that the model
can nevertheless illuminate generic aspects
of a candidate underlying theory.

When only one of the model's graviphotons, 
$F^{\mu\nu}$, is excited,
the bosonic part of our model
is given by
\bea
\ka \cl_{\rm sg}
&=&
-\frac 1 2 \sqrt{g} R
+\sqrt{g} ({\prt_\mu A\prt^\mu A + \prt_\mu B\prt^\mu B})/{4B^2}
\nonumber\\
&&
\qquad\!
-\frac 1 4 \ka \sqrt{g} M F_{\mu\nu} F^{\mu\nu}
-\frac 1 4 \ka \sqrt{g} N F_{\mu\nu} \tilde{F}^{\mu\nu}\; ,
\label{lag2}
\eea
where $M$ and $N$ are known functions of the scalars $A$ and $B$, 
$g=-\det (g_{\mu\nu})$, and $\tilde{F}^{\mu\nu}=\ve^{\mu\nu\rh\si}F_{\rh\si}/2$.
We can rescale
$F^{\mu\nu}\to F^{\mu\nu}/\sqrt{\ka}$
removing the explicit appearance 
of the gravitational coupling $\ka$
from the equations of motion.
We represent the model's fermions 
by the energy--momentum tensor of dust
$T_{\mu\nu} = \rh\,u_\mu u_\nu$
modeling, e.g., galaxies.
Here,
$u^\mu$ is a unit timelike vector
and $\rh$ is the fermionic energy density.
At tree level,
the fermionic matter is uncoupled from the scalars,
so that $T_{\mu\nu}$
is conserved separately.

With the phenomenological input of an isotropic homogeneous
flat Friedmann--Robertson--Walker Universe, 
we can take $F^{\mu\nu}=0$ on large scales.
Our cosmology then obeys
the Einstein equations
and the equations of motion for the scalars $A$ and $B$.
These equations can be solved analytically 
yielding a nontrivial dependence of $A=A_{\rm b}(t)$ and $B=B_{\rm b}(t)$ 
on the comoving time $t$. 
Consider now small localized excitations of $F_{\mu\nu}$
in the scalar background $A_{\rm b}$ and $B_{\rm b}$.
The effective Lagrangian $\cl_{\rm cosm}$ for such situations
in local inertial coordinates follows from Eq.\ \rf{lag2} and is
\beq
\cl_{\rm cosm}=
-\frac 1 4 M_{\rm b}F_{\mu\nu} F^{\mu\nu}
-\frac 1 4 N_{\rm b} F_{\mu\nu} \tilde{F}^{\mu\nu}\; ,
\label{efflag}
\eeq
where $A_{\rm b}(t)$ and $B_{\rm b}(t)$ 
imply the time dependence of $M_{\rm b}$ and $N_{\rm b}$. 
Comparison with the usual Maxwell Lagrangian $\cl_{\rm em} =
-\fr{1}{4 e^2} F_{\mu\nu}F^{\mu\nu}
- \fr{\th}{16\pi^2} F_{\mu\nu} \tilde{F}^{\mu\nu}$
establishes 
that $e^2 \equiv 1/M_{\rm b}$ and  $\th \equiv 4\pi^2 N_{\rm b}$.
Thus, 
$e$ and $\th$
acquire time dependencies via the varying background
$A_{\rm b}$ and $B_{\rm b}$.

The time dependence of $e$ is an interesting topic in itself, 
but in the present context 
the goal is to obtain 
the Lorentz- and CPT-violating Chern--Simons term $(k_{AF})^{\mu}A^{\nu}\tilde{F}_{\mu\nu}$ 
contained in the SME.
This can be achieved at the level of the action
via an integration by parts of the $\theta$-angle term. 
This establishes the desired result
\beq
\label{csterm}
\cl_{\rm cosm}\supset \frac 1 2 (\partial_{\mu}N_{\rm b})\, A_{\nu} \tilde{F}^{\mu\nu}\; .
\eeq
It is thus apparent 
that, 
starting from a Lorentz-invariant model, 
our supergravity cosmology 
has indeed generated one particular SME operator.

\section{Symmetry counting in the MCS model}
\label{symmetries}

In addition to the usual ten Poincar\'e invariances 
(four translations, three rotations, and three boosts), 
conventional electrodynamics 
possesses five further spacetime symmetries: 
one dilatation 
and four special conformal transformations. 
The inclusion of our Chern--Simons term $(k_{AF})^\alpha\, A^\beta\, \tilde{F}_{\alpha\beta}$ 
preserves translation invariance, 
since $(k_{AF})^{\mu}$ is assumed to be constant in the minimal SME. 
However, 
$(k_{AF})^{\mu}$ has mass dimensions 
suggesting 
that dilatation and conformal symmetry are violated.   
One further expects 
the Lorentz group to be broken down to the appropriate three-dimensional little group 
associated with $(k_{AF})^{\mu}$. 
This suggests 
that the MCS model maintains seven spacetime symmetries---four translations 
and three (of the original six) Lorentz invariances. 
This section argues 
that there is, in fact, 
one additional spacetime symmetry
if $(k_{AF})^{\mu}$ is lightlike: 
a combination of a boost and a dilatation. 

We begin by streamlining our notation $(k_{AF})^{\mu}\to k^{\mu}=(k,\vec{k})$
and by recalling 
that a dilatation,
also called a scale transformation, 
takes 
$A^{\mu}(x)\to e^{\rho}A^{\mu}(e^{\rho}x)$, 
where the size of the dilatation is determined 
by the parameter $\rho$. 
A dilatation therefore takes 
\beq
\label{dil}
\int\! d^4 x\,{\mathcal L}_{\rm MCS} \to \int\! d^4 x\, ({\mathcal L}_{\rm M}+e^{-\rho}\,{\mathcal L}_{\rm CS}) \neq \int\! d^4 x\,{\mathcal L}_{\rm MCS}\;, 
\eeq
where a suitable change of integration variables is understood.
It is apparent 
that the conventional piece ${\mathcal L}_{\rm M}$ 
and the Chern--Simons extension ${\mathcal L}_{\rm CS}$ transform differently. 
Moreover, 
the difference between the original and the transformed actions 
is not a boundary term 
establishing the non-invariance of ${\mathcal L}_{\rm MCS}$ under dilatations. 

We next look at Lorentz transformations, 
which can be implemented via $\Lambda^{\mu}{}_{\nu}(\vec{\theta},\vec{\beta})$. 
Here, 
$\vec{\theta}$ and $\vec{\beta}$ characterize rotations and boosts, respectively. 
Under such transformations, 
the MCS Lagrangian changes according to 
${\mathcal L}_{\rm MCS}=
{\mathcal L}_{\rm M}+{\mathcal L}_{\rm CS} \to {\mathcal L}_{\rm M}+ 
\Lambda^{\mu}{}_{\gamma}(-\vec{\theta},-\vec{\beta})\,k^{\gamma}\,A^{\nu}\,\tilde{F}_{\mu\nu}$. 
We have again suppressed the the dummy integration variables $x$. 
Next, we consider a special boost:  
$\vec{\beta}=\beta\hat{k}$, 
where $\hat{k}\equiv\vec{k}/|\vec{k}|$. 
Such a boost scales $k^{\mu}$ by a factor of $e^\beta$, 
so that 
$\Lambda^{\mu}{}_{\gamma}(\vec{0} ,-\beta\hat{k})\,k^{\gamma}\,A^{\nu}\,\tilde{F}_{\mu\nu}=\exp(\beta)\, k^\mu\, A^\nu\, \tilde{F}_{\mu\nu}\neq{\mathcal L}_{\rm CS}$
yielding 
\beq
\label{LT}
{\mathcal L}_{\rm MCS} \to {\mathcal L}_{\rm M}+e^{\beta}\,{\mathcal L}_{\rm CS} \neq {\mathcal L}\;. 
\eeq
This shows 
that symmetry under boosts along $\hat{k}$ is broken, 
as expected. 

Although each individual transformation \rf{dil} and \rf{LT} 
no longer determines a symmetry, 
the specific form of these transformations shows 
that a dilatation {\it combined} with 
a suitable boost along the spatial direction of a lightlike $k^{\mu}$ 
remains an invariance of ${\mathcal L}_{\rm MCS}$. 
We can verify this explicitly 
by studying the currents 
\beq
\label{Dcurrent}
D^{\mu}\equiv\theta^{\mu\nu}x_{\nu}
\eeq 
and 
\beq
\label{Lcurrent}
J^{\mu}_{\alpha\beta}\equiv \theta^{\mu}_{\alpha}x_{\beta}-\theta^{\mu}_{\beta}x_{\alpha}\;.
\eeq 
These are the usual dilatation and Lorentz currents, respectivly. 
To extract from Eq.\ \rf{Lcurrent} 
the current corresponding to a boost along $\hat{k}$, 
we decompose $k^{\mu}$ into its purely timelike 
and its purely spacelike part 
$k^{\mu}=k\,(k_T^{\mu}+k_S^{\mu})$, 
where $k_T^{\mu}=(1,\vec{0})$ and $k_S^{\mu}=(0,\hat{k})$. 
The desired current component 
is then given by $J^{\mu}_{\alpha\beta}\,k_S^{\alpha}\,k_T^{\beta}$. 
The divergences of these currents 
satisfy 
\beq
\label{Ddiv}
\partial_{\mu}D^{\mu}=-{\mathcal L}_{\rm CS}
\eeq
and 
\beq
\label{Ldiv}
\partial_{\mu}J^{\mu}_{\alpha\beta}\,k_S^{\alpha}\,k_T^{\beta}=+{\mathcal L}_{\rm CS}\;. 
\eeq 
It again becomes clear 
that $D^{\mu}$ and $J^{\mu}_{\alpha\beta}\,k_S^{\alpha}\,k_T^{\beta}$ 
are not conserved individually. 
However, 
their sum $Q^{\mu}\equiv D^{\mu}+J^{\mu}_{\alpha\beta}\,k_S^{\alpha}\,k_T^{\beta}$ 
is, in fact, conserved.
An explicit gauge-invariant expression for $Q^{\mu}$ can be obtained:\cite{hariton07} 
\beq
\label{explexpr}
Q^{\mu}=\big[\quar\, \eta^{\mu}_{\nu}F^2+F^{\mu\alpha}F_{\alpha\nu}\big]\big[x^{\nu}+(k_T\!\cdot\! x)\,k_S^{\nu}-(k_S\!\cdot\! x)\,k_T^{\nu}\big]\;.
\eeq
We can thus see 
that in the lightlike MCS model 
an additional conserved current 
relative to the spacelike and timelike cases
exists. 
With Killing-vector techniques, 
one can show 
that this is the only additional conservation law in the present context.\cite{hariton07}
This extended symmetry structure 
is described by the Lie algebra sim(2).\cite{hariton07}

\section{Vacuum Cherenkov radiation in the MCS model}
\label{cherenkov}

The Lorentz- and CPT-violating SME coefficients 
act in many respects like a background. 
This analogy is particularly well suited 
for the electrodynamics sector of the SME
because this sector exhibits many parallels 
to the conventional Maxwell theory in macroscopic media. 
It is therefore natural to ask 
as to whether such analogies can be exploited 
to identify possible phenomenological Lorentz- and CPT-breaking effects 
in the SME 
that can be employed for tests.

One conventional effect inside a macroscopic medium is 
that the phase speed of light $c_{ph}$ 
can be slowed down relative to the vacuum $c>c_{ph}$. 
It then becomes possible 
for ordinary charges $q$ to travel faster than light inside this medium: 
\beq
\label{cherenkovcondition}
v_q>c_{ph}\equiv\fr{\omega}{|\vec{p}|}
\eeq
Here, $v_q$ is the charge's speed, 
$\omega$ the photon frequency, 
and $\vec{p}$ the photon wave vector. 
It turns out 
this configuration is unstable 
in the sense 
that these fast charges are decelerated rapidly
through the emission of photons. 
This well established effect is called Cherenkov radiation. 
It can be employed, for example, 
in modern particle detectors. 
 
Can the Cherenkov effect  
also occur in our MCS model? 
This is indeed the case, 
which can be seen as follows. 
One can verify\cite{cherenkov} 
that Condition \rf{cherenkovcondition} 
continues to hold in the presence of Lorentz and CPT violation.
It follows 
that the MCS plane-wave dispersion relation $\omega=\omega(\vec{p})$ 
must be investigated. 
This dispersion relation 
is determined by 
\beq
\label{CSdr}
p^4+4\,k^2p^2-4(k\cdot p)^2=0\;,
\eeq
where $p^{\mu}=(\omega,\vec{p})$. 
For a given $\vec{p}$, 
Eq.~\rf{CSdr} determines a quartic equation in $\omega$, 
so that there are four branches of solutions. 
Two of these branches lie inside the momentum-space lightcone 
where $\omega(\vec{p})>\vec{p}$, 
which is inconsistent with Requirement \rf{cherenkovcondition}. 
However, 
the other two branches are located outside the lightcone 
where $\omega(\vec{p})<\vec{p}$ 
and the Cherenkov condition \rf{cherenkovcondition} is satisfied. 
We conclude 
that the Cherenkov effect can indeed occur in vacuum 
within the context of the MCS model. 

An important criterion for experimental Lorentz tests
is the rate for vacuum Cherenkov radiation. 
Within the MCS model, 
a classical calculation 
treating the charge as an external source 
yields\cite{cherenkov}
\beq
\label{Chrate}
\dot{P}^{\mu}=-\textrm{sgn}(k_0)\,\fr{q^2}{4\pi}\fr{k_0^4}{{\vec{k}}^2}\,(0,\hat{k})
\eeq
in the charge's rest frame. 
Here, 
$\dot{P}^{\mu}$ is the rate of four-momentum radiation 
and $q$ the particle's charge. 
It is apparent 
that the rate is suppressed: 
it is second order in the SME coefficient $k^{\mu}$. 
In the MCS model, 
vacuum Cherenkov radiation is 
therefore phenomenologically less interesting. 


\end{document}